**Title:** Hyper-lithography

**Authors Information**

Authors: Jingbo Sun,[1] Tianboyu Xu[1] and Natalia M. Litchinitser[1]*

Affiliations:

[1] Department of Electrical Engineering, The State University of New York at Buffalo, Buffalo, NY 14260, USA

*Correspondence and requests for materials should be addressed to : natashal@buffalo.edu

**The future success of integrated circuits (IC) technology relies on the continuing miniaturization of the feature size, allowing more components per chip and higher speed[1-3]. Extreme anisotropy opens new opportunities for spatial pattern compression from the micro- to nano-scale[4]. Such compression, enabling visible light-based lithographic patterning not restricted by the fundamental diffraction limit, if realized, may address the ever-increasing demand of IC industry for inexpensive, all-optical nanoscale lithography. By exploiting strongly anisotropic optical properties of engineered nanostructures, we realize the first experimental demonstration of hyperlens-based photolithography, facilitating optical patterning below the diffraction limit using a diffraction-limited mask. We demonstrate that the diffraction-limited features on a mask can be de-magnified to form the subwavelength patterns on the photoresist using visible light. This unique functionality, enabled by the hyperbolic dispersive properties of the medium combined with the cylindrical shape of the structure, opens a new approach to the future all-optical nanolithography.**

The remarkable developments in the IC industry are largely driven by the continuing progress in photolithography. Photolithography is the most widely used fabrication technique in the integrated circuit industry, where it is almost an exclusive way of pattern transfer from masks into semiconductor wafers. As the number of components on integrated circuit chip increases, further advances in feature size reduction on the nanoscale are needed. Some of the conventional approaches to the reduction of the feature size rely on decreasing the wavelength of light used in photolithography[5-9] or increasing the numerical aperture of the objective[10-11]. However, it has shown increasingly challenging in technique and expensive in fabrication as feature sizes decrease, since the diffraction still sets an ultimate limit in the optical resolution of conventional optical systems at about a half wavelength of the used light[12]. The diffraction limit originates from the fact that evanescent waves, carrying the high frequency information that defines the smallest features of the object, decay very fast in the conventional optical materials at far field. Other approaches to nanoscale optical lithography include near-field optical microscopy-based nanolithography[13], evanescent near-field optical lithography[14], surface-plasmon interference nanolithography[15], and various nanoscale aperture-based optical nano-patterning[16-19]. However, in a majority of them, either the pattern size is directly related to the feature size on the mask or nanoscale features are recorded in a step-by-step, scanning manner. The emergence of hyperbolic metamaterials might lead to a true paradigm shift in optical nanolithography. These materials are unique in that they enable an efficient transmission of evanescent waves carrying the information about the smallest features of the object[4, 20]. This remarkable property is facilitated by extreme anisotropy of hyperbolic materials[20,21]. Moreover, when shaped into a cylinder or a sphere, hyperbolic metamaterials form a so-called hyperlens that not only preserves the evanescent waves but also converts them into propagating waves and vice versa[22-25]. Therefore, we are able to overcome the diffraction limit by applying the hyperlens to the photolithography system and obtain fine nano patterns on the photoresist even using visible light and masks with the diffraction-limited features.

To realize the de-magnifying nanoscale photolithography system, we consider an extremely anisotropic, cylindrically shaped metamaterial consisting of alternating layers of metal and dielectric, as shown in Fig. 1. In microscopy, the sub-wavelength information carried by evanescent wave components from the sub-wavelength objects on the inner surface of the hyperlens will be magnified to the scale above the diffraction limit while it is propagating along the radial direction which can be finally converted into the propagating waves[22-24]. On the other hand, using the reciprocity theorem, the hyperlens may be used in a reverse way such that the incident light enters on the outer surface containing a macroscopic mask and are collected on the inner surface of the device. In this case, the structure may function as a de-magnifier that spatially compresses the features on the mask to the nanoscale[26-30]. Indeed, as the waves propagate along the radial direction toward the center of the hyperlens, tangential wave vectors increase so that the waves can propagate in the hyperlens. As a result, a micro- (or macro-) scopic pattern of the mask can be de-magnified to a sub-diffraction-limited pattern on the inner surface of the hyperlens. In contrast to the conventional photolithography, where the size of the lithographically-produced pattern is a function of the feature size on the mask and the numerical aperture of the lens, here the feature size is only related to the inner/outer radius of the de-magnifying hyperlens, which allows sub-wavelength, nanoscale patterns using micro- or even, in principle, macroscale masks.

In our photolithography system,, a de-magnifying hyperlens was fabricated between the Cr mask and the photoresist, as shown in Fig. 1. The hyperbolic metamaterial consisting of alternating metal and dielectric layers was designed such that it possesses a very large negative dielectric permittivity in the direction normal to the layers (radial direction) and positive dielectric permittivity less than 1 along the layers (tangential direction) at the working wavelength. As a result of such extreme anisotropy, the corresponding equi-frequency contour is a very flat hyperbolic shape, enabling efficient microscopic pattern compression to the sub-wavelength scale by converting and preserving the sub-wavelength information carried by the large components of the wave vector. Consequently, the de-magnified pattern with all wavelength details is imprinted exactly onto the photoresist covering the inner surface of the hyperlens. In this approach, a nanoscale pattern can be realized using a conventional macroscale mask and visible light wavelength, eliminating the need for the development of UV-specific optical components, new resists, and nanoscale masks.

An experimentally realized de-magnifying hyperlens consists of 13 layers of alternating layers of Ag and $Ti_3O_5$. A 1:1 filling ratio was chosen along with a 30 nm thickness for each layer [24]. The permittivities of the $Ti_3O_5$ and Ag at the wavelength of 405nm are 5.85 and -4.84+0.22i, respectively. According to the Maxwell-Garnett theory, dielectric permittivity tensor components of the multilayered structure can be determined by the following:

$$\varepsilon_\theta = \varepsilon_d f + \varepsilon_m (1-f), \tag{1a}$$

$$\varepsilon_r = \frac{\varepsilon_d \varepsilon_m}{\varepsilon_m f + \varepsilon_d (1-f)}. \tag{1b}$$

Using Eq. (1), we found that the permittivity along the layers was $\varepsilon_\theta$ =0.51+0.11$i$ and the one perpendicular to the layers was $\varepsilon_r = -53+14.1i$.

Propagation of the incident 405nm-light inside the de-magnifying hyperlens was modeled using a finite element-based method implemented in COMSOL Multiphysics (see supplementary information). The letter "U" with a line-width of 300nm was used as the mask pattern inscribed on the outside of the hyperlens. The light is polarized orthogonal to cylindrical axis of the hyperlens. The simulation results show the intensity profile cross-sections at different distances from the mask inside the hyperlens and the image in the vertical plane perpendicular to the cylindrical groove surface. These results show the evolution of the beams from the slits on the mask until they reach the inner surface of the hyperlens. The nearly flat equi-frequency contour of the designed hyperbolic material ensures that the beams beyond the mask are well confined along the radial direction and are gradually compressed while still being resolvable inside the hyperbolic material. Comparing the bottom (initial) and top (final) images, we conclude that the initial microscale U-shape pattern inscribed on the mask is reproduced on the resist but with the sub-wavelength features size, as shown in Fig. 2(a). In order to prove that the sub-wavelength compression is enabled by the de-magnifying hyperlens, we replaced the multilayer with a conventional $MgF_2$ lens of same shape and size, as shown in Fig. 2(b). It can be seen that the beam transmitted through the slits on the mask diffracts very fast and results in a strongly distorted pattern on the inner side of the lens.

The proposed device was fabricated by depositing 13 alternating layers of Ag(30nm) and $Ti_3O_5$(30nm) onto a cylindrical groove in a glass substrate. Before the deposition, a layer of 50nm Cr was deposited on glass and the "UB" pattern with a line width of 300nm was milled using a focused ion beam, as shown in Fig. 3 and Fig. 4(a). A thin layer of polymethyl methacrylate (PMMA) (60nm) was filled into the patterns by spin coating to form a smooth surface for the hyperlens deposition. Following the multilayer deposition, a 50 nm photoresist layer (diluted S1805, see supplementary information) was deposited on the inner surface of the hyperlens to record the de-magnified pattern from the hyperlens. The sample was illuminated from the glass substrate/Cr mask side with a 405nm laser for 15s. Next, it was immersed in the developer solution MF-26A for 40s to develop the pattern on the photoresist. Note that the feature size of initial pattern on the Cr mask is above the diffraction limit for this incident light wavelength.

Figure 4(b) shows scanning electron microscope (SEM) image of the compressed (sub-wavelength) UB pattern appearing on the photoresist layer. The widths of all vertical lines are around 170nm confirming that the original pattern was de-magnified by approximately 1.8 times. In contrast, a reference sample consisting of a $MgF_2$ cylindrical lens with the same thickness as a hyperlens was fabricated on top of the mask and then spin coated with the photoresist. Figure 4(c) shows that after the same exposure and develop process, a strongly blurred image was produced lacking the sub-wavelength details on the inner surface of the $MgF_2$ lens. Indeed, the photoresist exposed by light transmitted through the two arms of the "U" and the small circle in "B" letter are completely removed due to the strong diffraction of light inside the $MgF_2$. The large circle in "B" still appeared but with much wider line width, about 500nm.

In summary, we experimentally demonstrated de-magnifying cylindrical hyperlens. The original patterns with a feature size of 300nm were scaled down to 170nm, which is below the diffraction limit for wavelength of the incident light beam. Such performance was made possible owing to the unique ability of the hyperlens to support propagation of the evanescent waves. The de-magnification factor can be further increased by increasing the total thickness of the hyperlens. Higher de-magnification factors would further relax the requirements on the maximum feature-size of the pattern recorded on the mask. Furthermore, isotropic de-magnification can be facilitated by a spherically-shaped hyperlens. While theoretically, the resolution of the hyperlens made of uniform hyperbolic medium is only determined by the geometrical parameters of the hyperlens and therefore, is unlimited, material losses, details of a particular fabrication process, and quantum optics effects may need to be taken into account to estimate the lower limit of practically achievable resolution. This first experimental demonstration of a de-magnifying hyperlens may give rise to an entirely new branch of optical lithography.

**Acknowledgments:** The authors acknowledges support of this work by the US Department of Energy Award DE-SC0014485.

We thank the help of Dr. Bangzhi Liu from Pennsylvania State University in the materials deposition.

**Authors contributions:** J. S and N. M. L. proposed the idea developed in this work. J. S. made the design and did the simulation. J. S. and T. X. did the fabrication and the characterization of the sample. N. M. L. and J. S. wrote the paper . N. M. L. supervised this work.

**Competing financial interests**

The authors declare no competing financial interests.

**Figure legends**

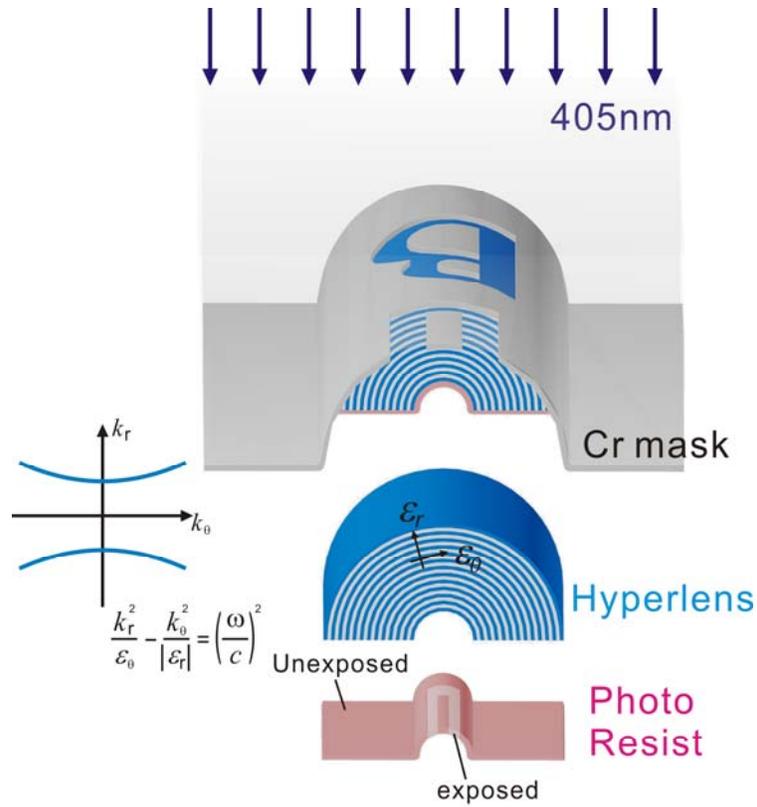

**Fig. 1.** Design of de-magnifying hyperlens-assisted optical lithography. Schematic of the lithography system, consisting of a metal/dielectric hyperlens fabricated between the Cr mask and a photoresist layer. Light illuminating the mask propagates through the anisotropic multilayer structure of the hyperlens and exposes the photoresist. The inset shows flat hyperbolic equi-frequency contours, corresponding to the designed hyperbolic metamaterials, which enables spatial compression of the microscale pattern inscribed on the mask into the nanoscale pattern recorded on the photoresist.

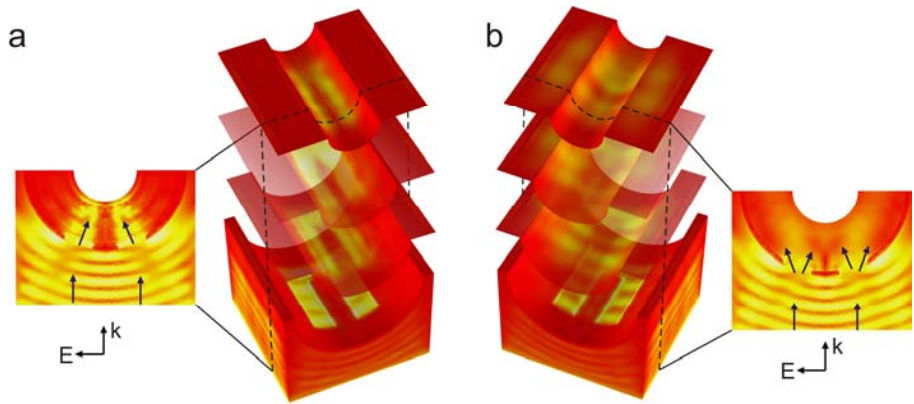

**Fig. 2.** Numerical simulations of the light propagation through the de-magnifying hyperlens (a) and through the conventional MgF$_2$ lens of the same shape (b). The pattern on the mask is a "U" shape slit. (a) An intensity profile of a compressed U-shape appears on the inner surface of the hyperlens. The cross-section cutting at the two arms of the U

shows that beams past the mask are well confined along the radial direction in the hyperlens material even when it is compressed to the sub-wavelength size. (b) A blurred image appears on the inner surface of the conventional MgF$_2$ lens. The cross section illustrates that beams beyond the mask diffract and interfere with each other, resulting in significant image distortion/feature size broadening.

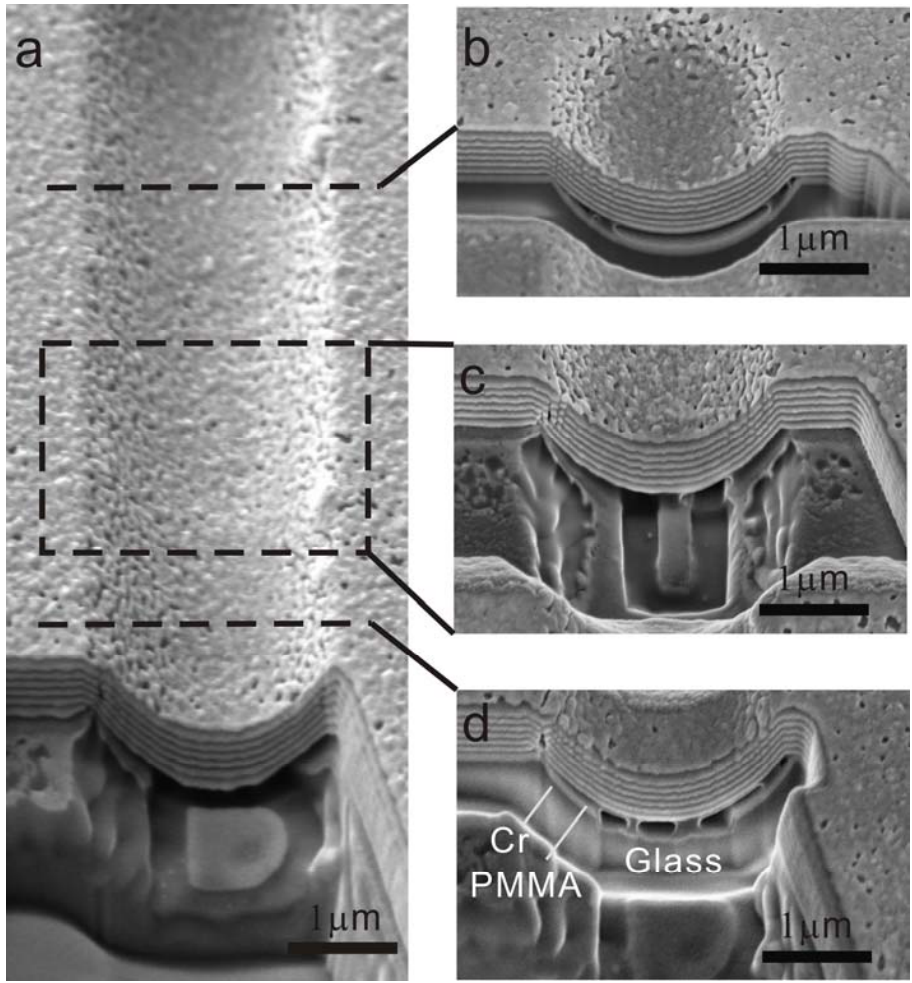

**Fig. 3.** The scanning electron microscope images of the de-magnifying hyperlens cross-sections. (a) Hyperlens on the Cr mask with "UB" pattern; (b) Cross-section view cut through the hyperlens corresponding to the unpatterned region; (c) "U" shape pattern on the mask under the hyperlens; (d) Cross-section view cut through the hyperlens at upper part of the letter "B". From the substrate to the top layer: glass substrate, PMMA, Cr mask and multilayered hyperlens structure.

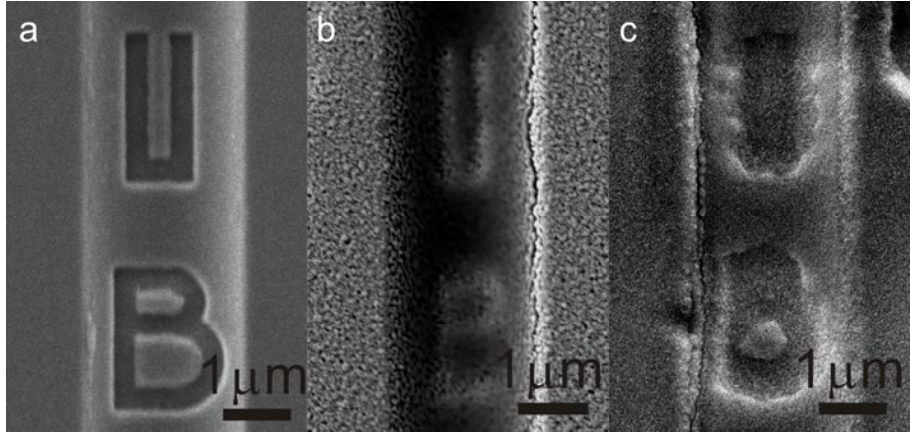

**Fig. 4.** Initial mask and lithographically recorded patterns. (a) Original "UB" pattern on the Cr mask. (b) Sub-wavelength pattern on the photoresist recorded using hyperlens. (c) Lithographically recorded pattern using reference sample consisting of MgF$_2$ lens.

**Supplementary Materials:**

Materials and Methods

Material properties and numerical simulation

Optical characterization

**Supplementary Materials:**

**Materials and Methods:**

In this section, we outline the main fabrication steps of the de-magnifying hyperlens. First, a 100 nm Cr film was deposited on a flat glass substrate using the electron beam evaporation method. This layer was used as the etching mask. Next, the focused ion beam system (FIB, Carl Zeiss AURIGA) was implemented to mill a 50-nm-wide and 8-μm-long slit in the Cr film. A cylindrical groove was obtained through using Cr film as a mask as well as isotropic wet etching in a buffered oxide etch (in a 6:1 ratio).

In the next step, Cr film was removed using CR-7 Cr etchant, and then a 50nm Cr was deposited by electron beam evaporation as the photomask. The FIB was used to mill the pattern "UB" on the Cr layer at the bottom of the cylindrical groove. The width of the lines in this pattern was 300nm. Subsequently, 60nm thick PMMA A2 (MicroChem) layer was spin-coated on the Cr film to fill the pattern and form a smooth surface underneath the hyperlens.

Next, the hyperlens was formed by depositing alternating Ag and Ti$_3$O$_5$ thin layers (7 layers of Ag and 6 layers of Ti$_3$O$_5$) on top of the Cr photomask. Finally, the Shipley1805 (MicroChem) photoresist diluted with propylene glycol monomethyl ether acetate (MicroChem) with 20% (v/v) was spin coated on the hyperlens to record the de-magnified image from the hyperlens.

In parallel, a dielectric lens was also prepared by depositing 390nm thick MgF$_2$ layer on top of the photomask instead of the multilayered metamaterial. Then, the same photo resist was spin coated on to this dielectric lens as a reference sample.

**Material properties and numerical simulation:**

Light propagation through the de-magnifying hyperlens was investigated in detail using finite-element method implemented in COMSOL Multiphysics™ 5.1. The structure was composed of glass substrate, Cr ($\varepsilon = -10+12i$), PMMA ($\varepsilon = 2.25$), Ag ($\varepsilon = -4.84+0.22i$)/Ti$_3$O$_5$ ($\varepsilon = 5.85$) multilayer and S1805 ($\varepsilon = 2.25$) layer. The Cr film contained a "U"-shaped pattern, where the slit width of 300nm was filled with PMMA in agreement with the experimentally developed samples. We also performed numerical simulations for the reference sample. In this case, the multilayered structure was replaced by MgF$_2$ layer whose refractive index is 1.38 at 405 nm.

**Optical characterization:**

The device was characterized using a 405 nm, 20mW laser source. First, the laser beam was focused using a 10× objective and then collimated using a lens with a 50mm focusing length. The orientation of the sample was such that the axis of the cylindrical trench was orthogonal to the direction of polarization of the laser beam.

The device was exposed for 15 seconds and then the sample was developed using the MF-26A developer for 40 seconds. After the development, the de-magnified image of the original letters "UB" appeared on the inner surface of the hyperlens. The resulting de-magnified line-width was measured to be 170nm.

In order to assure that identical exposure condition have been applied to both hyperlens and reference sample, for the case of the reference sample the incident beam was first attenuated to the level of that in the hyperlens case. In order to achieve the same attenuation level, we used 13 layers Ag/Ti$_3$O$_5$ flat structure that was deposited together with the hyperlens as an attenuator. In this way, it guaranteed that the photo resist in the reference sample was exposed in the similar intensity with the hyperlens sample. Then, the sample was developed by MF-26A for 40 seconds to get the SEM image shown in Fig. 4(c).